\documentclass[10pt,conference]{IEEEtran}
\IEEEoverridecommandlockouts

\usepackage{hyperref}
\hypersetup{
    colorlinks=true,
    linkcolor=blue,
    urlcolor=cyan,
    citecolor=blue,
}

\usepackage{amsfonts}
\usepackage{algorithmic}
\usepackage{graphicx}
\usepackage{textcomp}
\usepackage{xcolor}
\usepackage{verbatim}
\usepackage{listings}
\usepackage{multirow}
\usepackage[caption=false]{subfig}
\usepackage{enumitem}
\usepackage{framed}
\usepackage{colortbl}
\usepackage{soul}
\usepackage{booktabs}
\usepackage{balance}

\usepackage[numbers,sort&compress]{natbib} 

\definecolor{lightgray}{gray}{0.95}
\definecolor{darkblue}{rgb}{0,0,0.5}
\definecolor{lightred}{rgb}{1, 0.8, 0.8}
\definecolor{lightyellow}{rgb}{1, 1, 0.8}
\definecolor{lightgreen}{rgb}{0.6, 1, 0.6}
\definecolor{mint}{rgb}{0.74, 0.99, 0.79}
\definecolor{forestgreen}{rgb}{0.13, 0.55, 0.13}

\def\BibTeX{{\rm B\kern-.05em{\sc i\kern-.025em b}\kern-.08em
    T\kern-.1667em\lower.7ex\hbox{E}\kern-.125emX}}

\begin{document}

\title{OSS License Identification at Scale: A Comprehensive Dataset Using World of Code
\thanks{Replication package available at: \url{https://zenodo.org/records/14279932}}
}

\author{
\IEEEauthorblockN{Mahmoud Jahanshahi, David Reid, Adam McDaniel, Audris Mockus}
\IEEEauthorblockA{\textit{Department of Electrical Engineering and Computer Science} \\
\textit{University of Tennessee, Knoxville, USA} \\
\{mjahansh, dreid6, amcdan23, audris\}@utk.edu}
}

\maketitle

\begin{abstract}

  The proliferation of open source software (OSS) and different
  types of reuse has made it incredibly difficult to perform an
  essential legal and compliance task of accurate license
  identification within the software supply chain.  This study
  presents a reusable and comprehensive dataset of OSS licenses,
  created using the World of Code (WoC) infrastructure.  By scanning
  all files containing ``license'' in their file paths, and applying
  the approximate matching via winnowing algorithm to identify the
  most similar license from the SPDX and Open Source list, we found and identified
  5.5 million distinct license blobs in OSS projects.  The dataset
  includes a detailed project-to-license (P2L) map with commit
  timestamps, enabling dynamic analysis of license adoption and
  changes over time.  To verify the accuracy of the dataset we use
  stratified sampling and manual review, achieving a final accuracy
  of 92.08\%, with precision of 87.14\%, recall of 95.45\%, and an
  F1 score of 91.11\%. This dataset is intended to support a range
  of research and practical tasks, including the detection of
  license noncompliance, the investigations of license changes, study
  of licensing trends, and the development of compliance tools. The
  dataset is open, providing a valuable resource for
  developers, researchers, and legal professionals in the OSS
  community.

\end{abstract}

\begin{IEEEkeywords}
Software Licenses, Open Source Software, Open Source Licenses, World of Code
\end{IEEEkeywords}

\section{Introduction}\label{intro}

As the open-source software (OSS) ecosystem has expanded rapidly, 
it has given rise to a diverse array of projects, each characterized 
by different licenses and licensing practices. A fundamental value of 
OSS lies in the ability to reuse code, either through dependency management
or by directly copying and potentially maintaining (vendoring) it. 
Many licenses impose specific requirements on code usage, such as 
the obligation to publish derived works under GPL licenses. The 
reuse supply chains are often complex and difficult to trace. 
Consequently, accurately identifying OSS licenses across the entire
supply chain is crucial for understanding the legal frameworks
that govern OSS distribution and use.
Such understanding is crucial for ensuring license compliance, 
fostering collaboration, and mitigating risks within software 
supply chains. Despite the significance of OSS licensing, existing 
studies often fall short of covering the entire supply chain by 
focusing on specific ecosystems, subsets of projects, or lack 
essential attributes needed to identify timing and project 
information. Without this information, it becomes impossible 
to reconstruct the dynamics or pinpoint the location of 
licenses within the supply chain~\cite{reid2023applying}.

This work makes a step in addressing these challenges by compiling
a reusable and comprehensive dataset of OSS licenses.
To accomplish that we exploit the World of Code
(WoC)~\cite{ma2021world} that contains version history from a nearly
complete collection of publicly accessible software projects.  
We start from all files that contain ``license'' in their file
paths and discover over 10M blobs (distinct strings) associated with
these files. For each we then find the most similar license from several 
``official'' license collections. To accomplish that we use
winnowing algorithm, a fingerprinting technique known for its
ability to match text with minor variations, such as differences in
formatting, even in cases where the text is embedded or has
undergone slight modifications~\cite{serafini2022efficient}. 
Our method successfully identifies and maps over 5.5 million distinct
license blobs to known licenses, generating a project-to-license (P2L) 
map enriched with commit timestamps. 
Furthermore, we enhance our dataset by incorporating the previously
published dataset by~\citet{gonzalez2023software}.

This dataset fills critical gaps in the study of OSS licensing by
providing: 1) a large-scale, cross-platform resource for analyzing
license adoption, change, evolution, and compliance, 2) dynamic tracking
capabilities through commit timestamps, enabling longitudinal
studies of licensing practices, and 3) a foundation for developing
tools and methods to address challenges in OSS license compliance
and compatibility.

The dataset and its associated methodology have been designed with
reusability and scalability in mind, ensuring that it can be readily
adopted by researchers, practitioners, and legal professionals.
By making the dataset openly available, we aim to foster new
research in software engineering and contribute to better practices
in the OSS ecosystem. 

\section{Related Work and Contributions}

Understanding OSS licensing practices has been the focus of numerous studies, ranging from license identification to compliance analysis.
These studies have contributed valuable insights but are often limited in scope, scale, or methodology.

\subsection{Comprehensive Identification of License Blobs}  

Previous studies like~\citet{wu2024large} and~\citet{xu2023lidetector} focus on explicit license declarations in metadata files, while others, such as~\citet{feng2019open}, use static analysis to detect embedded license texts in binaries. 
While these methods and datasets advance license text identification, they do not address partial matches or embedded license texts, which are common in OSS projects.

In contrast, our work leverages the winnowing algorithm, a robust fingerprinting method, to identify both partial and full matches of license blobs across millions of files, even when license texts are embedded or slightly modified. 
This approach enhances precision and ensures comprehensive identification, capturing both standard and non-standard licensing practices in OSS repositories.

\subsection{Broad Scale and Scope of Analysis}  

Prior studies have often been limited in scope, focusing on specific platforms or datasets.
Large-scale efforts have identified license files but overlooked contextual information, such as project associations or temporal data.
For example, \citet{zacchiroli2022large} introduced a dataset of 6.5 million blob-license text variant tuples (spanning 4.3 million unique blobs), enabling analyses of text diversity and NLP-based modeling of license corpora.
However, their work focuses on cataloging text variants rather than linking licenses to their usage within projects.
Similarly, \citet{gonzalez2023software} documented 6.9 million blob-license tuples (representing 4.9 million unique blobs), but the emphasis remained on cataloging rather than exploring connections to broader supply chain dynamics.

Our study expands the scope of previous research by analyzing the entire OSS landscape through the World of Code (WoC) infrastructure.
We match over 5.5 million license blobs to known licenses and map them to specific OSS projects and their histories.
This comprehensive project-to-license (P2L) mapping facilitates detailed tracking of licensing practices across platforms, bridging the gap between text-level variability and actionable project-level insights.

\section{Methodology}\label{method}

\subsection{World of Code Infrastructure}

World of Code (WoC)\footnote{\url{https://worldofcode.org}} is an infrastructure designed to cross-reference source code changes across the entire OSS community, enabling sampling, measurement, and analysis across software ecosystems~\cite{ma2019world,ma2021world}.
It functions as a software analysis pipeline, handling data discovery, retrieval, storage, updates, and transformations for downstream tasks~\cite{ma2021world}.

WoC offers maps connecting git objects and metadata (e.g., commits, blobs, authors) and higher-level maps like project-to-author connections, author aliasing~\cite{fry2020dataset}, and project deforking~\cite{mockus2020complete}.
We use WoC to identify all license blobs and their associated projects\footnote{Version V, latest at the time of this study.}, employing the concept of deforked projects~\cite{mockus2020complete} to avoid biases from forks and duplicates.

\subsection{License Blob Identification}

We start by using the blob-to-filepath maps (b2f) in WoC to list all filepaths for each blob, specifically searching for those with ``license'' in their filepath.
Using blob hashes ensures that any license blob, even if associated with a ``license'' filepath in only a single project, will still be identified.
Using blob-to-project maps (b2P), we then identify all projects containing that blob, which means that we do not require the blob to have the ``license'' filepath in every project.
This ensures high recall in detecting potential license-related blobs by leveraging the collective metadata of public repositories.
This approach resulted in over 10 million distinct potential license blobs.

Since there are relatively few known licenses, we anticipate that most of these blobs are similar licenses with minor variations, such as differences in whitespace, formatting, or non-essential information.
The main challenge is matching these varied license blobs to known licenses.

We use licenses from the Open Source Initiative\footnote{\url{https://github.com/OpenSourceOrg/licenses}} and the Software Package Data Exchange (SPDX)\footnote{\url{https://github.com/spdx/license-list-data}}, which include 103 and 635 licenses, respectively.
To match the 10 million potential license blobs with these known licenses, we apply winnowing, an efficient local fingerprinting algorithm~\cite{schleimer2003winnowing}.

Winnowing is a document fingerprinting technique often used in plagiarism detection.
It generates fingerprints by sliding a window over hashed words in a document and selecting the smallest hash value in each window.
This reduces the data needed for document representation, enabling faster and more memory-efficient comparisons while maintaining accuracy.

Using winnowing, we matched over 7 million potential license blobs to one of the known licenses (see Table~\ref{tbl:score}).
We assess the reliability of these matches by calculating a matching score, defined as the number of shared winnowing signatures divided by the total winnowing signatures between two files.
This score, as shown in Equation \ref{eq:win}, measures the similarity between the potential license blob and the known license, helping to verify the match's accuracy.

\begin{equation}
S = \frac{c(A \cap B)}{c(A \cup B)} 
\label{eq:win}
\end{equation}

\begin{description}
    \item[$S$:] Matching score.
    \item[$A$:] Set of signatures in document A.
    \item[$B$:] Set of signatures in document B.
    \item[$c(X)$:] Count function for the number of elements in set $X$.
\end{description}

\begin{table}[ht]
\centering
\caption{Potential License Blobs Matching Scores}
\resizebox{0.99\linewidth}{!}{
\begin{tabular}{l|rrr}
    \toprule
    & \textbf{Count} & \textbf{Percentage} & \textbf{Percentage} \\ 
    & & (Relative to) & (Overall) \\ 
    \midrule
    Potential Blobs & 10,093,268 & 100\% & 100\% \\ 
    Winnowing & 9,794,559 & 97\% (Potential Blobs) & 97\% \\
    Matched & 7,167,046 & 73.2\% (Winnowing) & 71\% \\
    \multicolumn{4}{c}{\dotfill} \\
    $S <= 0.2$ & 795,532 & 11.1\% (Matched) & 7.9\% \\
    $0.2 < S <= 0.4$ & 239,091 & 3.3\% (Matched) & 2.4\% \\
    $0.4 < S <= 0.6$ & 264,667 & 3.7\% (Matched) & 2.6\% \\
    $0.6 < S <= 0.8$ & 435,283 & 6.1\% (Matched) & 4.3\% \\
    $0.8 < S <= 1$ & 5,432,473 & 75.8\% (Matched) & 53.8\% \\
   \bottomrule
\end{tabular}}
\label{tbl:score}
\end{table}

We categorized matching scores into five groups: below 20\%, 20-40\%, 40-60\%, 60-80\%, and above 80\%.
As shown in Table~\ref{tbl:score}, 97\% of blobs generated winnowing signatures.
We randomly sampled 30 blobs from the 3\% that did not and manually confirmed they had no meaningful content.
Of the 9.7 million blobs, 73\% matched a known license (sharing at least one winnowing signature), with 75\% of these matches scoring above 80\%.

To assess match reliability, we sampled 20 blobs from each score group and manually compared them to the known license using `diff`.
Given the manual nature of the verification process, choosing 20 samples for each bucket provides a manageable workload while still offering a sufficient range of data to detect patterns and inconsistencies.
Our investigation revealed that matches in buckets with scores below 80\% were not reliable enough, showing meaningful differences.

We then focused on scores above 80\% and conducted another stratified sampling based on score range (80-85, 85-90, 90-95, 95-100) and the number of signatures (above/below 100).
In each group, 20 matches were sampled.
The differences fell into three main categories: 1) identical content with different formatting, 2) identical content with non-license text, and 3) identical content with additional clauses.

The second category was acceptable, as we do not claim a blob contains only the matched license.
However, the third, with additional clauses, was concerning as it could alter the license’s nature. Detailed results are in Table~\ref{tbl:score-sample}.

\begin{table}[ht]
\centering
\caption{Matching Score Samples}
\resizebox{0.99\linewidth}{!}{
\begin{tabular}{ccr|ccc}
    \toprule
    \textbf{Signatures} & \textbf{Score} & \textbf{Total Count (\%)} & \textbf{Gr. 1} & \textbf{Gr. 2} & \textbf{Gr. 3} \\ 
    \midrule
    \multirow{4}{*}{\textbf{$<=100$}} & 
    80-85 & 85,294 (1.6\%) & 17 & 3 & 0 \\ 
    & 85-90 & 150,046 (2.8\%) & 17 & 3 & 0 \\
    & 90-95 & 197,875 (3.6\%) & 20 & 0 & 0 \\
    & 95-100 & 4,502,264 (82.9\%) & 20 & 0 & 0 \\
    \multicolumn{6}{c}{\dotfill} \\
    \multirow{4}{*}{\textbf{$>100$}} &
    80-85 & 67,235 (1.2\%) & 10 & 9 & 1 \\ 
    & 85-90 & 52,894 (1\%) & 17 & 2 & 1 \\
    & 90-95 & 60,583 (1.1\%) & 18 & 2 & 0 \\
    & 95-100 & 316,282 (5.8\%) & 20 & 0 & 0 \\
   \bottomrule
\end{tabular}}
\label{tbl:score-sample}
\end{table}

We observed only two mismatches: one in the 80-85\% range and one in the 85-90\% range (both in the over 100 signatures group).
Based on this, we determined that setting the threshold at 85\% ensures reliable license identification.
Above this threshold, critical mismatches---where additional clauses could alter the license---are extremely rare.
Since over 90\% of identified blobs had fewer than 100 winnowing signatures, the 85\% threshold balances comprehensiveness and precision, capturing most valid matches while minimizing misleading results.
This approach aligns with prior research emphasizing high similarity thresholds to reduce false positives in textual matching (e.g., \cite{kapitsaki2017automating}).
As a result, 5,294,666 distinct blobs were matched with a known license.

For the remaining 2.5 million potential blobs with no matches, we randomly sampled 30 and manually investigated them.
Only 5 contained license-related content, either mentioning a license name or linking to a license URL.
The other 25 were unrelated to licenses.

\subsection{Project to License Mapping}

\begin{figure*}
\centering
\resizebox{0.85\linewidth}{9.8cm}{
\includegraphics{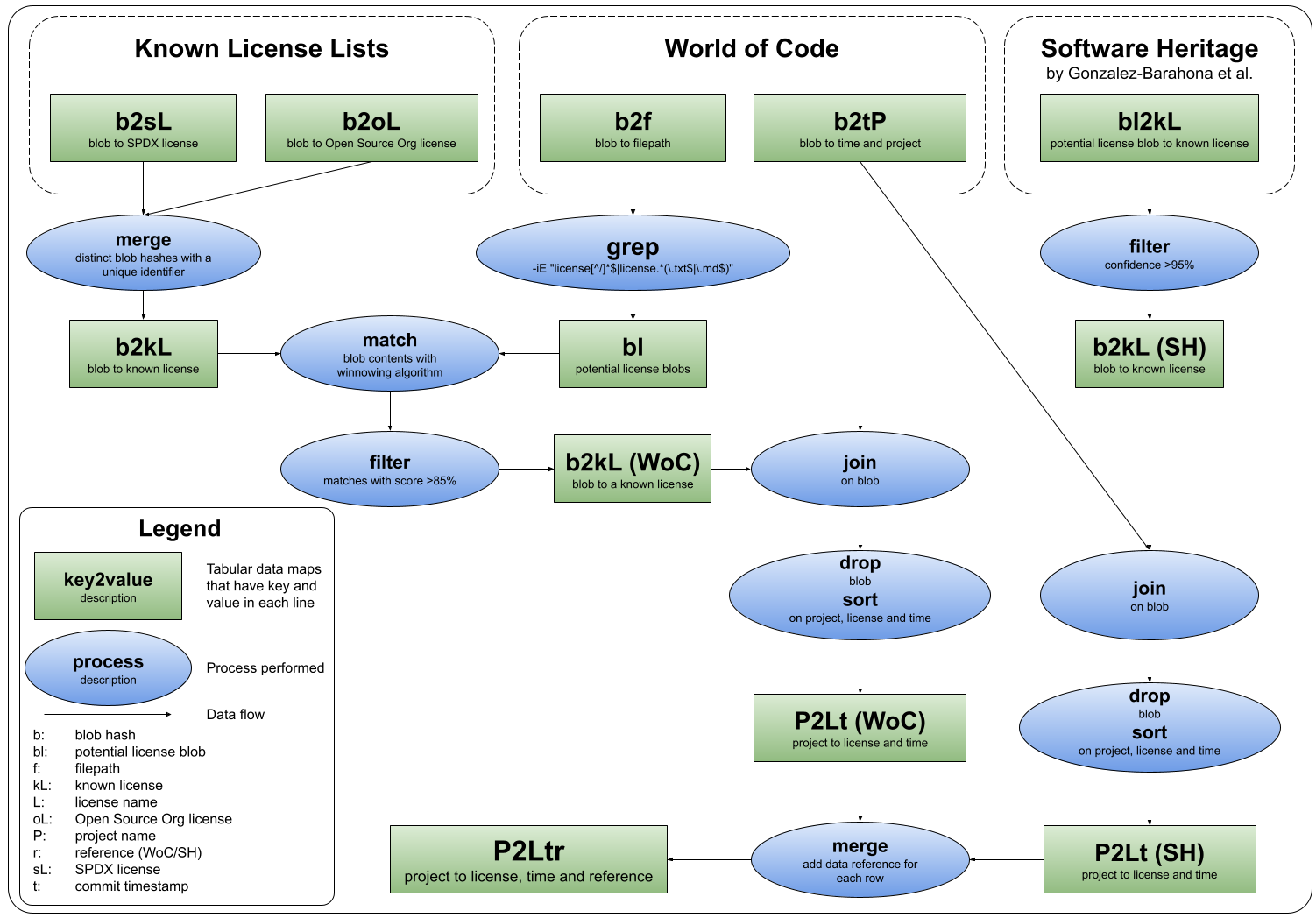}}
\caption{License Identification Data Flow Diagram}
\label{fig:dfd}
\end{figure*}

To create the project-to-license (P2L) map, we use the 5.5 million matched license blobs and join them with WoC's blob-to-time project (b2tP) map, which links blobs to the projects they were committed to, along with commit timestamps.
This produces a table mapping each project to a known license and the time of the commit (see Figure~\ref{fig:dfd}).

However, a blob’s presence in a project’s latest status cannot be confirmed solely from commit history, as it might have been removed later.
To address this, we use the project-to-last-commit (P2lc) and tree-to-objects (t2all) maps from WoC.
The P2lc map links projects to their last commit at the time of the latest WoC update (Version V), allowing us to retrieve the list of all blobs in a project’s current state by joining P2lc, c2dat (commit-to-tree), and t2all maps.
This method not only provides all the times at which a blob was committed to a project but also verifies whether it still exists in the project.

The final table is saved as a semicolon-separated file containing three fields\footnote{
For more information on accessing this data, please visit \url{https://github.com/woc-hack/tutorial}}:

$Project\_ID;License;Commit\_Time$

The $Commit\_Time$ field is in the ``YYYY-MM'' format and represents the commit timestamp when the license blob was committed to the project.
This field may also have an ``invalid'' value, indicating that the commit timestamp was not valid (e.g., a future time due to discrepancies in the user's system time).
Additionally, if the license blob was found in the latest status of a project, the time is ``latest''.

\subsection{P2L Verification}

For the Project-to-License (P2L) verification, we initially sampled 1,000 projects from approximately 130 million to evaluate the effectiveness of our license assignment methodology.
This sample size was chosen to provide a statistically significant subset for manual verification while balancing the need for reliability with the practical constraints of manual inspection.

We stratified the sample into three groups: 
1) Projects with matched licenses, where our automated process successfully matched license blobs to known licenses, 
2) Projects with license blobs but no matched licenses, where license blobs were identified, but no matching known license could be confirmed, and 
3) Projects without any license blobs, where no license blobs were detected during the automated search.

This sampling approach was designed to cover a wide range of license detection scenarios, ensuring a comprehensive evaluation.
Graduate students manually reviewed the sampled projects as part of a class assignment, focusing on verifying the license information.
Out of the 1,000 sampled projects, we received meaningful responses for 580 projects, distributed as follows: 291 with matched licenses, 139 with license blobs but no matches, and 150 without any license blobs.
The results are presented in Table~\ref{tbl:confusion}.

\begin{table}[ht]
\centering
\caption{License Detection Confusion Matrix Across Stages}
\label{tbl:confusion}
\resizebox{0.99\linewidth}{!}{%
\begin{tabular}{l|cc|cc|cc}
    \toprule
    \textbf{Stage} & \multicolumn{2}{c|}{\textbf{Initial}} & \multicolumn{2}{c|}{\textbf{Adjusted}} & \multicolumn{2}{c}{\textbf{Refined}} \\
    \midrule
    & \textbf{License} & \textbf{No License} & \textbf{License} & \textbf{No License} & \textbf{License} & \textbf{No License} \\
    \midrule
    \textbf{Matched} & 210 & 81 & 210 & 31 & 210 & 31 \\
    \textbf{Not Matched} & 22 & 267 & 22 & 267 & 10 & 267 \\
    \midrule
    \textbf{Accuracy} & \multicolumn{2}{c|}{82.24\%} & \multicolumn{2}{c|}{90.00\%} & \multicolumn{2}{c}{92.08\%} \\
    \textbf{Precision} & \multicolumn{2}{c|}{72.16\%} & \multicolumn{2}{c|}{87.14\%} & \multicolumn{2}{c}{87.14\%} \\
    \textbf{Recall} & \multicolumn{2}{c|}{90.52\%} & \multicolumn{2}{c|}{90.52\%} & \multicolumn{2}{c}{95.45\%} \\
    \textbf{F1 Score} & \multicolumn{2}{c|}{80.31\%} & \multicolumn{2}{c|}{88.79\%} & \multicolumn{2}{c}{91.11\%} \\
    \bottomrule
\end{tabular}}
\end{table}

Our license detection method demonstrated reasonable performance with an initial accuracy of 82.24\%, precision of 72.16\%, recall of 90.52\%, and an F1 score of 80.31\%.

However, several factors must be considered when interpreting these results:
first, of the 81 projects identified as having matched licenses, 39 no longer exist on GitHub, preventing license verification, and
second, in 11 projects, the license was absent in the latest status, which does not necessarily indicate a false positive, as the license could have been removed after an earlier commit.
After excluding these cases, we are left with 31 false positives.
Adjusting for these, our revised performance metrics show significant improvement: accuracy increases to 90.00\%, precision to 87.14\%, recall remains at 90.52\%, and the F1 score rises to 88.79\%.

For the 22 false negatives (where licenses were not detected), further investigation revealed that only 10 had a missed license blob, which was matched but fell slightly below our 85\% threshold.
The remaining 12 projects only referenced a license (e.g., in the README) without including the actual license file in the repository, so they were not expected to be matched by our method.
By excluding these 12 false negatives, which fall outside our method's intended scope, we can more accurately assess its performance.
The recalculated metrics show an accuracy of 92.08\%, precision of 87.14\%, recall of 95.45\%, and an F1 score of 91.11\% (see Table~\ref{tbl:confusion}).

\subsection{Complementing Data}

Although our P2L map already demonstrated strong performance in manual verification, we incorporated the previously published dataset by~\citet{gonzalez2023software} to enhance data comprehensiveness.
Their dataset includes only blobs and their detected licenses using ScanCode~\cite{scancode_toolkit}. 
We filtered data to blobs with license detection confidence 95\% or higher and applied the same process described earlier to map these blobs to commits and projects, enabling us to determine the time and project in which each license was committed.
The merged table (see Figure~\ref{fig:dfd}) includes a column indicating the license detection method for each entry: either our method (1-WoC) or the Software Heritage dataset method (2-SH)~\cite{gonzalez2023software}.

\section{Applications}

The dataset described in this work provides a robust foundation for addressing key challenges in open source software (OSS) licensing.
Below, we discuss use cases supported by the dataset and illustrate them with examples from ongoing research conducted by the authors, which are currently under review and cannot be cited directly.

\subsection{Ensuring License Compliance}

Managing license compliance is a critical issue in OSS, where licensing conflicts or noncompliance can lead to significant legal and ethical challenges.
This dataset enables research into understanding and mitigating compliance risks.
For instance, the dataset has been used to analyze how licensing conflicts arise from code reuse across OSS projects.
These insights underscore the need for advanced compliance tools that leverage comprehensive project-to-license mappings to detect and address potential license violations.

\subsection{Analyzing Licensing Trends and Practices}

Understanding how OSS licenses are selected and evolve over time is essential for improving licensing practices and fostering innovation.
The dataset supports large-scale analyses of license adoption trends, revealing patterns and the factors influencing license choices (e.g.~\citet{vendome2017license}).
For example, it has been used by the authors to explore the dynamics of license adoption, examining the role of social, technical, and ideological factors in shaping these decisions.
The dataset's extensive coverage allows researchers to track the evolution of licenses within and across OSS ecosystems, providing actionable insights for developers and policy-makers.

\subsection{Supporting Ecosystem Studies and Tool Development}

The dataset’s comprehensive project-to-license mapping has broad applicability in supporting ecosystem studies and tool development. 
Such applications include investigating how licensing practices influence collaboration and innovation in OSS communities, enabling the creation of automated tools for license verification, detecting noncompliance, recommending suitable licenses, and providing a resource for educating developers on licensing implications and best practices.

\section{Limitations}

\paragraph{Scope of License Identification}

The current methodology focuses on files explicitly named ``license'' or located in license-related directories, which may miss license information embedded in source code headers, build scripts, or files with unconventional names.
These gaps particularly affect older or unconventional OSS projects.
Expanding the search scope and using natural language processing (NLP) or pattern recognition could improve coverage. 
To partially address this, we incorporate the dataset by \citet{gonzalez2023software} to enhance comprehensiveness.

\paragraph{Implicit Licensing Practices}

Implicit licensing practices, such as referencing licenses by name or URL in README files or documentation, are not captured, potentially leaving gaps for permissively licensed projects.
Future work could parse these files to link references to known licenses.

\paragraph{Data Completeness and Noise}

Finally, while robust heuristics minimize errors, some non-license files may be misidentified, and legitimate licenses in non-standard formats could be excluded.
Feedback mechanisms and automated quality checks could further enhance reliability.

\section*{Acknowledgments}
This work was supported in part by the National Science Foundation under Award Numbers 1901102 and 2120429.

\balance
\small
\bibliographystyle{IEEEtranN}
\setlength{\bibsep}{0pt plus 0.3ex} 
\bibliography{ref}

\end{document}